\documentclass[prd,twocolumn]{revtex4}
\usepackage{graphicx}
\usepackage{amsmath,amssymb,amsthm,amsbsy}
\usepackage{latexsym}
\usepackage{amsfonts}
\usepackage{amssymb}

\renewcommand{\nu}{v}

\begin{document}

\title{Path Integral Over Black Hole Fluctuations}

\author{James W. York, Jr.}
\affiliation{Department of Physics,
Cornell University, Ithaca, New York 14853}
\email{york@astro.cornell.edu, bss28@cornell.edu}

\author{Bjoern S. Schmekel}
\affiliation{Department of Physics,
Cornell University, Ithaca, New York 14853}
\email{bss28@cornell.edu}

\begin{abstract}
Evaluating a functional integral exactly over a subset of metrics
that represent the quantum fluctuations of the horizon of a
black hole, we obtain a Schroedinger equation in null
coordinate time for the key component of the metric. The
equation yields a current that preserves probability if
we use the most natural choice of functional measure. 
This establishes the existence of blurred horizons and a
thermal atmosphere. It has been argued previously that the
existence of a thermal atmosphere is a direct concomitant of the thermal
radiation of black holes when the temperature of the hole is greater
than that of its larger environment, which we take as zero.
\end{abstract}

\maketitle

\section{Introduction}
Hawking has shown that black holes emit radiation that has the
expectation values of thermal (``gray body'') radiation \cite{hawking1975}.
Wald sharpened the result by removing the restriction
to expectation values \cite{wald1975} and DeWitt \cite{dewitt1975} gave a complete
analysis of all the modes of radiation, including those that
``hug'' the region $2M < r \le 3M$.
York \cite{york1983} argued in detail that there must be a thermal atmosphere
surrounding the hole as a result of the zero-point
fluctuations of the metric of the horizon: $\Delta g \sim \hbar ^{1/2}$.
A further argument was made to the effect that this thermal atmosphere
was not a phenomenon different from the Hawking radiation, but is its
physical concomitant. This was inferred from a careful estimation
of its temperature. The result was within two percent of the Hawking
temperature for the lowest quasi-normal oscillation modes of the hole,
which were assumed to be the frequencies at which zero-point
fluctuations would naturally occur. 
In his analysis the matter fields were completely neglected. Vacuum
fluctuations break the degeneracy of the event horizon, the apparent
horizon and the time-like limit surface and radiation can be emitted.
One expects that the temperature of the emitter can only depend on the
Schwarzschild radius $r_s$, $\hbar$ and possibly $G$. Every combination thereof
with unit temperature (in units where the Boltzmann constant is equal to $1$)
has the form $G^{n-1} \hbar ^n r_s^{1-2n}$. Since the radiation in York's
picture is due to vacuum zero-point fluctuations the temperature is independent 
of $G$, too. Therefore, the temperature must be of the order of $\hbar / r_s$ in
both York's and Hawking's picture. Now we actually know which result quantum
gravity has to yield.

A physical argument was given as
to why the apparent ultraviolet divergences associated with high frequency modes
do not actually occur. Alternatively and equivalently, the textbook definition
of a black hole has the locus of its horizon defined by null geodesics - 
infinite frequency massless particles. These, of course, do not actually exist
because there is an associated width for any real ``photon'' or wave packet.

Except for the lowest frequency waves for a given masslesss spin, the outgoing waves do not
propagate freely. This is well known in
classical studies of quasi-normal waves. Higher frequencies are trapped by the hole
as zero-mass, speed of light ``moons'' completely surrounding and ``hugging'' the hole
and therefore raising no tides and creating no observable effects. The lowest frequency modes,
however, do create a splitting of the event horizon, the apparent horizon and the timelike limit
surface. The ``blurred'' region of horizons is the thermal atmosphere.

The point of view in \cite{york1983} was similar to a way that one can
view Welton's treatment \cite{welton} of the Lamb shift: A Dirac hydrogen
atom does come into equilibrium with the zero-point fluctuations of the 
electromagnetic field. The latter break a degeneracy. This shift in energy
levels is the Lamb shift and it shows how the hydrogen atom comes into
equilibrium by means of this adjustment in its zero temperature vacuum. 
In \cite{york1983} it is argued that a black hole cannot do this when
bathed in zero temperature gravitational metric fluctuations (``virtual gravitons'').
It must form a thermal atmosphere and radiate. This is a form of the
fluctuation-dissipation phenomenon which will be considered in detail elsewhere.
This fluctuation-dissipation effect does not restore equilibrium
because the surrounding zero-temperature is not the black hole temperature.
In the present case, the dissipation is a gradual ``run-away'' further and
further from equilibrium as the black hole shrinks and becomes hotter - unless 
the hole somehow finds itself in an equally hot region of space.

In this paper we improve the semi-classical arguments in \cite{york1983} by
a path integral method. In this context we wish to mention techniques in
loop quantum gravity \cite{meissner2004}, anomaly cancellation \cite{wilczek2005} 
and string theory \cite{strominger1996}. However, these approaches 
take the black hole somehow as a fixed ``background'' feature, which it cannot be.
The advantage of the approach presented here
is its mathematical simplicity and its connection to a powerful physical
picture. Only very few simplifying assumptions are necessary to obtain
a Schroedinger equation. However, we are by no means claiming that our treatment is
the final word, or that we have gotten to the very bottom of the phenomenon. 


\section{Classical Equations of Motion}
A spherical black hole can be described by the metric
\begin{eqnarray}
\nonumber
ds^2 = & - & F(\nu,r,\theta) d \nu^2 + 2 d \nu dr 
\\
& + & \psi^2(r) \left ( d \theta^2 + \sin^2 \theta d \phi^2 \right )
\label{metric}
\end{eqnarray}
where $F$ is a function of $\nu$, $r$, and $\theta$ and $\psi$ is a function of $r$.
This resembles the kind of metric used by Bardeen \cite{bardeen1981} and, later, York \cite{york1984}
to study the backreaction of black hole evaporation.
Surfaces of constant $\nu$ are null surfaces whose null tangents and null normals are $\partial / \partial r$.
All deviations from the classical path are taken to be of this form which
simplifies our treatment substantially. In particular it is very easy to avoid ghosts.
Like the treatment of the one-dimensional harmonic oscillator in non-relativistic
quantum mechanics this truncated ``superspace'' quantization freezes degrees of freedom
the system actually posseses. This freezing, of course, violates the uncertainty relation for those degrees 
of freedom. Nevertheless, the one-dimensional harmonic oscillator is still
useful even though the ground state energy is too small by a factor of three.
However, the nonlinear nature of general relativity makes it hard to predict
what the exact effect of fixing some of the dynamical fields will be (cf. \cite{kuchar1989}).
We will comment on this complication later.

Classically, we expect in the absence of matter $\psi(r)=r$ and $F(\nu,r,\theta) = 1 - \frac{2m(\nu,\theta)}{r}$.
Since the commutator in the semi-classical approach \cite{york1983} was evaluated at fixed ``time'' $\nu$
and orbital ``quantum number'' $l$ (i.e., no transitions between levels with different $l$ were considered, a simplification)
we keep the $\nu$ and $\theta$ dependent part fixed. Thus, $m(\nu,\theta)=M$ where $M$ is the average mass of the 
hole which we will assume in the remainder of the paper. Varying only the $r$-dependent
part of $F(\nu,r,\theta)$ the Euler-Lagrange equation for $F$ will turn out to leave $F$ undetermined
because the Lagrangian depends only linearly on $F$ and its derivatives. Varying $\psi(r)$ alone and fixing
$F$ at its classical value $F(\nu,r,\theta) = 1 - \frac{2m}{r}$ gives a Lagrangian which does not
look manifestly local in ``time'', i.e. the Lagrangian depends on $r$ explicitly. Thus, we decide to vary either $\psi(r)$ and/or the 
$r$-dependent part of $F$ and relate the two fields by the classically expected relation 
$F = 1 - \frac{2m}{\psi(r)}$. This will allow us to deal with one scalar field which is a function of one variable only. 
The Hilbert action is 
\begin{eqnarray}
S=\int dr \int_{\nu_0}^{\nu_0 + \frac{2 \pi}{\omega}} d \nu \int_0^{2\pi} d \phi \int_0^{\pi} d \theta \mathcal{L}
\label{action}
\end{eqnarray}
with the Lagrangian 
\begin{eqnarray}
\mathcal{L} = \sin \theta \left [  \psi^2 F_{,rr} + 4\psi \psi_{,r} F_{,r} + 4\psi \psi_{,rr} F + 2\psi_{,r}^2 F - 2 \right ] .
\end{eqnarray}
We integrate over one period $2\pi/\omega$ of quasi-normal mode oscillations. Since $\mathcal{L}$ is independent 
of $\nu$ this only results in an additional normalization factor. 
Note that the factor $\sqrt{-g}=\psi^2(r) \sin \theta$ is already included
in the Lagrangian. We verify that our approach is meaningful at the classical level.
Varying the Lagrangian with respect to $F$ (which is just $-g_{\nu \nu}$) we can obtain the equations of 
motion from the Euler-Lagrange equation
\begin{eqnarray}
\frac{d^2}{dr^2} \frac{\partial \mathcal{L}}{\partial F_{,rr}} -
\frac{d}{dr} \frac{\partial \mathcal{L}}{\partial F_{,r}} +
\frac{\partial \mathcal{L}}{\partial F} = 0
\end{eqnarray}
In order to vary the three components of the metric tensor together
we have to express $\mathcal{L}$ in terms of $F$ and its derivatives
only, i.e. $\mathcal{L}=\mathcal{L}(F,F_{,r},F_{,rr})$. Assuming
$r \approx 2M$ and $(1-F)^{-1} \approx 1+F$ we have
\begin{eqnarray}
\mathcal{L} = \sin \theta \left [
2(1-F) - \frac{8m}{1-F} F_{,r} - \frac{4m^2}{(1-F)^2} F_{,rr} \right ]
\end{eqnarray}
and finally
\begin{eqnarray}
- r^{-1} F_{,r} - F_{,rr} - r^{-2}(1-F) =0
\end{eqnarray}
which is equivalent to the following linear combination of components of the
Einstein tensor
\begin{eqnarray}
G^{\nu}_{\nu} - G^{\theta}_{\theta} - G^{\phi}_{\phi} = 0
\end{eqnarray}
However, we prefer to vary the Lagrangian with respect to $\psi$,
relating $\psi$ and $F$ as before where $r_0 = 2M$.
This allows us to get rid of linearizations like $(1-F)^{-1} \approx 1 + F$
used above. 
The Euler-Lagrange equation for $\psi$ is simply 
\begin{eqnarray}
\psi_{,rr}=0~ 
\label{eomh}
\end{eqnarray}
whose solution is $\psi(r)=c_1 + c_2 r$ where $c_1$ and $c_2$ are arbitrary constants.
The constants are not an artifact of gauge freedom. They can be determined by the 
``boundary condition'' that equates the constant $m$ in the metric to the mass
of the black hole.

Having convinced ourselves that the classical equation of motion is correct
we now compute the action which we will need later for our path integral prescription.
Evaluating all integrals in Eq. \ref{action} except the $dr$-integral and using
$F=1-2m/\psi(r)$ we obtain for the action without any approximations
\begin{eqnarray}
S = \frac{16 \pi^2}{\omega} \int dr \left ( \psi_{,r}^2 - 3M \psi_{,rr} + 2\psi \psi_{,rr} - 1 \right )
\end{eqnarray}
In classical mechanics the Lagrangian can be multiplied by an arbitrary normalization constant
$k$. In quantum gravity however the normalization is not arbitrary. We define a new action
for which we explicitly retain the normalization constant $k$ which will be determined later.
Also, because in classical mechanics an arbitrary number can be added to the Lagrangian without
altering the equations of motion we have added a scalar number to our new action which cancels
the constant part in the parenthesis. The second term in the parenthesis is a pure boundary
term which we have omitted in our new action. Finally, the third term is integrated by parts
and the emerging boundary term is absorbed as well. Thus, our new redefined action is
\begin{eqnarray}
S = - k \frac{16 \pi^2}{\omega} \int dr \psi_{,r}^2
\label{finalaction}
\end{eqnarray}  
No linearizations were used to obtain Eq. \ref{finalaction}.
Also note that it is irrelevant whether we use $\psi$ or any other power
of $\psi$ as our field. The invariance of general relativity under diffeomorphisms
allows for any coordinate transformation we desire.
\section{Path Integral Evaluation}
Now we can compute the path integral for the $\psi(r)$ field
\begin{eqnarray}
\left <  \psi_f | e^{-iH \Delta r / \hbar} | \psi_i \right > = \int \mathcal{D} \psi(r)
e^{iS[\psi(r)]/ \hbar}~.
\label{pathintegral}
\end{eqnarray}
$\psi$ can take on values from minus to plus infinity. Despite the occurence of $\psi^2(r)$
in (\ref{metric}) every value of $\psi$ produces a different metric because of the constraint 
$-g_{\nu \nu} = F = 1 - 2m \psi^{-1}(r)$. If we had used $\psi(r)$ for $g_{\theta \theta}$
instead of $\psi^2(r)$ we would have had $-g_{\nu \nu} = F = 1 - 2m \psi^{-1/2}(r)$ and
integrating from minus to plus infinity (which simplifies the calculation) would result in complex values. 
Of course we can make use of a coordinate transformation (e.g. use $\psi^2$ instead of $\psi$ as the field) 
as long as we use the right functional measure.

To understand better the meaning of the metrics we sum over, let us look at the paths
$\psi = c_1 + c_2 r$. This is of course only a subset of metrics we sum over.
Because the Lagrangian only depends on the derivative of $\psi$ any constant $c_1$
added to it results in the same value of the action.
The amount of overcounting is independent of the non-constant part of $\psi$
and will be compensated by choosing an appropriate normalization factor for the action.
Summing over $\psi$ with different values for $c_2$ simply means we are adding vacuum
metrics corresponding classically to different physical masses.

Choosing an approriate measure is still an unresolved problem in quantum gravity 
(cf. \cite{anselmi1992} for proposed measures and associated problems like anomaly cancellation).
We use the following ansatz for the measure \cite{anselmi1992,dewitt1967} 
\begin{eqnarray}
\mathcal{D} \psi(r) = \prod_i C^{-1} \psi^n(r_i) d \psi(r_i)
\label{measure}
\end{eqnarray} 
where $n$ is some number which in general is neither positive nor integral and $C$ is a constant.
Note that this problem for $n=0$ is now analogous to deriving the one-dimensional Schroedinger equation
from the Lagrangian $\mathcal{L}_S = \frac{\dot x^2}{2 m_e}$ using the path integral
prescription \cite{peskin} if we use the following substitutions:
\begin{eqnarray}
\nonumber
t & \longrightarrow & r
\\
\nonumber
x & \longrightarrow & \psi
\\
2 m_e & \longrightarrow & - \omega (16 \pi^2 k)^{-1}~.
\end{eqnarray}
Neither an expansion into Feynman diagrams nor a Wick rotation to a Euclidean path integral is necessary.
Since all derivatives in the Lagrangian are taken with respect to $r$ meaningful 
results can be only expected when summing over paths parametrized by $r$ which are null, not timelike.
This issue is not new. It has been observed before by Schmekel for the truncated superspace quantization of several 
different problems \cite{schmekel}. In our treatment $r$ is an affine parameter along a null line. 
Such a quantization along a null curve is known in quantum field theory as an infinite momentum frame
quantization \cite{soper} which alters the set of Feynman rules.
The quantization along a null curve has a very pleasant effect. An observer who is at rest with respect 
to a black hole will detect an enormous redshift of the radiation emitted by an infalling object. 
In the infinite momentum frame this redshift is absent. All the nonlinearities which could be present in the equations
of motion vanish and we are left with quantizing a theory which is
linear in our chosen frame. This helps justify our truncated superspace approach. Also note that despite
having used linearizations no quantities were linearized in deriving (\ref{finalaction}).

With the substitutions mentioned above we obtain the following Schroedinger equation
\begin{eqnarray}
i \hbar \frac{\partial}{\partial r} \varphi = \hbar^2 k \frac{16 \pi^2}{\omega}
\frac{\partial^2 \varphi}{\partial \psi^2} \equiv \hat H \varphi
\end{eqnarray}
where $\varphi \equiv \varphi(r,\psi)$. From the Schroedinger equation an uncertainty relation 
for $\Delta(\psi)$ and $\Delta (m_e \psi_{,r})$ can
be derived.
\begin{eqnarray}
m_e \left < \Delta \psi \right > \left < \Delta \psi_{,r} \right > \ge \frac{\hbar}{2} 
\end{eqnarray}
Thus, with
\begin{eqnarray}
\psi = \sqrt{g_{\theta \theta}} = r_0 + \frac{\delta g_{\theta \theta}}{2 r_0} + \mathcal{O}(r_0^{-2})	
\label{linrootgthth}
\end{eqnarray}
we obtain
\begin{eqnarray}
\left < \Delta g_{\theta \theta} \right > \left < \Delta \frac{d}{dr} g_{\theta \theta} \right > 
\ge \frac{\hbar}{2} \frac{1}{(vol)}
\label{uncertaintyrel}
\end{eqnarray}
with an appropriate choice of the normalization constant $k$ and $(vol)=\frac{4}{3} \pi (2\pi \omega^{-1})^3$. 
This uncertainty relation was found and used to derive the Hawking temperature and entropy \cite{york1983}.
Furthermore, our Schroedinger equation conserves the probability current. Thus, information 
cannot be lost in the formation of a thermal atmosphere, which, in turn, leads to evaporation
if the hole is hotter than its surroundings, which we have taken as having zero temperature.
The evaporation process would be unitary in our example. We point out, however, that latter
possibility cannot be established by the present arguments alone.

In appendix B we consider the case $n=1$ in the measure (\ref{measure}). An additional
term appears in the Schroedinger equation which violates conservation of probability current.
The resulting equation for $\varphi$ is not local in the sense that the extra term contains $\psi$ which is an argument
of $\varphi$. Similar problems occur for different choices with $n \neq 0$. Therefore, this equation does not treat 
all measures equally, but favours the ultra-local measure. This leads us to 
the conclusion that the previous case with $n=0$ is appropriate for our problem if we use $\psi$ as our field.

Note that for different fields a different measure may be necessary. For instance if we had used $\psi^2$
as our field $n=1$ would be an appropriate choice. In this case the linearization in Eq. \ref{linrootgthth} 
would not be necessary to arrive at Eq. \ref{uncertaintyrel}.

\section{Black Hole Horizon Width}
We employ the uncertainty relation to estimate the width of the horizon.
The tensor momentum corresponding to $g_{\theta \theta}$ and the magnitude of the lowest order
metric fluctuations have been found before \cite{york1983}
\begin{eqnarray}
\Delta g_{\theta \theta ,r} = \frac{1}{64 \pi M} \Delta g_{\theta \theta}
\\
(\Delta g_{\theta \theta})^2 \sim \frac{3 \hbar }{\pi^3 M^2}
\end{eqnarray}
This gives with $\delta r = r - r_0$
\begin{eqnarray}
<\delta r> = \sqrt{\frac{3\hbar}{\pi^3}}
\end{eqnarray}
which not surprisingly is in the order of the Planck length independently of $M$.
This results agrees with York's findings which were based on purely physical reasoning.
However, note that the Hawking temperature, {\it which in our framework is now a result from
a theory of quantum gravity}, does not scale as the Planck scale because it is independent of $G$.
\section{Summary and Conclusions}
It was shown by quantizing a subset of metrics that
a meaningful uncertainty relation can be obtained which can be used
to describe the blurring of the event horizon and the thermal atmosphere, and that transitions
between different states are unitary - at least for the simplest choice of functional measure. 
However, we have reasons to believe that this is the appropriate measure,
because other measures produce Schroedinger equations which violate locality.
The authors are currently investigating how this path integral quantization 
along null coordinates can be generalized to more general problems. 

\acknowledgments

We thank the referee for valuable comments. JWY thanks the National Science Foundation for
support under grant number PHY-0216986.

\appendix

\section{Path Integral Quantization ($n=0$)}

We would like to quantize the Lagrangian $\mathcal{L}=b~\psi_{,r}^2$. Following \cite{peskin}
we compute the contribution from the last path-slice $\psi'=\psi(\lambda-\epsilon)$ with thickness $\epsilon$ to the probability amplitude, i.e.

\begin{widetext}

\begin{eqnarray}
\varphi(\psi_i,\psi_f,\lambda) = \int_{-\infty}^{\infty} \frac{d\psi'}{C} \exp \left [
\frac{ib}{\hbar}  \frac{(\psi_f-\psi')^2}{\epsilon} \right ]
\varphi(\psi_i,\psi',\lambda-\epsilon)
\end{eqnarray}
For the discretization the following substitutions were used
\begin{eqnarray}
\nonumber
\psi \longrightarrow \frac{\psi_f + \psi'}{2} 
\\
\psi_{,r} \longrightarrow \frac{\psi_f - \psi'}{\epsilon}
\end{eqnarray}
The last slice at $r=\lambda$ and the first slice are denoted by $\psi_f = \psi(\lambda)$ and $\psi_i=\psi(0)$, respectively.
Expanding in powers of $\psi'-\psi_f$ gives

\begin{eqnarray}
\varphi(\psi_i,\psi_f,\lambda) = \int_{-\infty}^{\infty} \frac{d\psi'}{C} \exp \left [
\frac{ib}{\epsilon \hbar} (\psi_f - \psi')^2 \right ]
\left [ 1 + (\psi'-\psi_f) \frac{\partial}{\partial \psi_f} + \frac{1}{2} (\psi'-\psi_f)^2 \frac{\partial^2}{\partial \psi_f^2} + ...\right]
\varphi(\psi_i,\psi_f,\lambda - \epsilon)
\end{eqnarray}

Evaluating the integral gives

\begin{eqnarray}
\varphi(\psi_i,\psi_f,\lambda) = \frac{\sqrt{\pi \epsilon \hbar}}{C \sqrt{-ib}} 
\left [ 1 + \frac{i \epsilon \hbar}{4b} 
\frac{\partial^2}{\partial \psi_f^2} \right ] \varphi(\psi_i,\psi_f,\lambda - \epsilon)
\end{eqnarray}

\end{widetext}

In order to obtain a meaningful equation the factor in front of the bracket has to be $1$ and $C$
is chosen accordingly. Finally, we obtain

\begin{eqnarray}
i\frac{\partial}{\partial r} \varphi =  -
\frac{\hbar}{4b} \frac{\partial^2}{\partial \psi^2}  \varphi
\end{eqnarray}

\section{Path Integral Quantization ($n=1$)}

\begin{widetext}

\begin{eqnarray}
\varphi(\psi_i,\psi_f,\lambda) = \int_{-\infty}^{\infty} \frac{d\psi'}{2C} (\psi_f+\psi') \exp \left [
\frac{ib}{\hbar}  \frac{(\psi_f-\psi')^2}{\epsilon} \right ]
\varphi(\psi_i,\psi',\lambda-\epsilon)
\end{eqnarray}

The last slice at $r=\lambda$ and the first slice are denoted by $\psi_f = \psi(\lambda)$ and $\psi_i=\psi(0)$, respectively.
Expanding in powers of $\psi'-\psi_f$ gives

\begin{eqnarray}
\nonumber
\varphi(\psi_i,\psi_f,\lambda) = \int_{-\infty}^{\infty} \frac{d\psi'}{2C} (\psi_f+\psi') \exp \left [
\frac{ib}{\epsilon \hbar} (\psi_f - \psi')^2 \right ]
\\ \times
\left [ 1 + (\psi'-\psi_f) \frac{\partial}{\partial \psi_f} + \frac{1}{2} (\psi'-\psi_f)^2 \frac{\partial^2}{\partial \psi_f^2} + ...\right]
\varphi(\psi_i,\psi_f,\lambda - \epsilon)
\end{eqnarray}

Evaluating the integral gives

\begin{eqnarray}
\varphi(\psi_i,\psi_f,\lambda) = \frac{\psi_f \sqrt{\pi \epsilon \hbar}}{C \sqrt{-ib}} 
\left [ 1 - \frac{i \epsilon \hbar}{4b~\psi_f} \frac{\partial}{\partial \psi_f} + \frac{i \epsilon \hbar}{4b} 
\frac{\partial^2}{\partial \psi_f^2} \right ] \varphi(\psi_i,\psi_f,\lambda - \epsilon)
\end{eqnarray}

\end{widetext}

In order to obtain a meaningful equation the factor in front of the bracket has to be $1$ and $C$
is chosen accordingly. Note that the normalization constant $C$ includes $\psi_f$, i.e.
starting from Eq. \ref{pathintegral} we would have to use the measure
\begin{eqnarray}
\mathcal{D} \psi = \prod_i \left ( \frac{\epsilon}{4} \psi_{,r}(r_i) + \frac{1}{2} \psi(r_i) \right )
\frac{\sqrt{\pi\epsilon\hbar}}{\sqrt{-ib}} \psi(r_i) d \psi(r_i)
\end{eqnarray} 
in order to obtain an evolution equation. Ignoring this peculiarity, we obtain

\begin{eqnarray}
i\frac{\partial}{\partial r} \varphi =  \frac{\hbar}{4b ~ \psi} \frac{\partial}{\partial \psi}  \varphi
-\frac{\hbar}{4b} \frac{\partial^2}{\partial \psi^2}  \varphi
\end{eqnarray}

Now we can show that the latter equation does not conserve the probability current.

\begin{eqnarray}
\frac{\partial}{\partial r} | \varphi |^2 = \overline{\frac{\partial}{\partial r} \varphi}
+ \overline{\varphi} \frac{\partial}{\partial r} \varphi =
\\
-\frac{i\hbar}{4b} \left ( \overline{\varphi_{,\psi}} \varphi - \overline{\varphi} \varphi_{,\psi} \right )_{,\psi}
+\frac{i\hbar}{4b} \frac{1}{\psi} \left ( \overline{\varphi_{,\psi}} \varphi - \overline{\varphi} \varphi_{,\psi} \right )
\\
\equiv - j_{,\psi} + \frac{1}{\psi} j \quad\quad
\end{eqnarray}
The dependence of $C$ on $\psi$ and $\psi_{,r}$, the violation of conservation of probability current, and
the fact that the resulting Schroedinger equation is not local in the sense that it does
not only depend on $\varphi$ and its derivatives but also on $\psi$ suggest that the functional
measure with $n=0$ is to be used. 

\pagebreak

\end{document}